\documentclass[twoside]{article}
\usepackage{fleqn,espcrc2}

\usepackage{graphicx}
\usepackage[figuresright]{rotating}

\newcommand{\AmS}{{\protect\the\textfont2
  A\kern-.1667em\lower.5ex\hbox{M}\kern-.125emS}}
\newcommand{\beeq}{\begin{equation}}
\newcommand{\eneq}{\end{equation}}
\newcommand{\beeqar}{\begin{eqnarray}}
\newcommand{\eneqar}{\end{eqnarray}}

\title{Numerical Simulations of $d=3$ $SU(2)$ LGT in the
Dual Formulation 
      }

\author{N.D. Hari Dass\address{Institute of Mathematical Sciences, 
        C.I.T Campus, Taramani, Chennai 600 113, INDIA}%
        }
       
\begin{document}

\begin{abstract}
We have developed the techniques necessary for a numerical simulation of
$d=3$ $SU(2)$ Lattice Gauge theories in the dual formulation as originally
developed by Anishetty et al \cite {anishetty}. These include updating
techniques that preserve the constrained configuration space, efficient
evaluation of 6-j symbols and a certain problem associated with the positive
indefiniteness of the weight factors.
\vspace{1pc}
\end{abstract}

\maketitle

\section{Motivation}
The outstanding problems in QCD are understanding the mechanisms of confinement,
chiral symmetry breaking and the spectrum of the theory including the vacuum
structure. Lattice gauge theories have gone a long way in shedding light on some
of these important issues. Nevertheless, we are still not in a position to, say,
discriminate between different mechanisms for confinement like the dual 
superconductor mechanism, the $Z_N$-fluxon mechanisms etc. In particular a
detailed picture of QCD in terms of the dual variables is still at a preliminary
level. Therefore it is desirable to probe alternative formulations of LGT that
place greater emphasis on the dual variables. 

The dual variables simplify the
description of some observables while complicating some others. In the formulation
to be discussed here, we shall see that while the t'Hooft disordered variables
become diagonal, the Wilson loop operators become quite complicated. The
continuum limit of LGT's is in the extreme weak coupling regime and here many
of the gauge invariant observables like plaquettes approach unity so in
extracting the scaling part of various correlation lengths large "perturbative"
corrections need to be subtracted. But in the dual formulation there are many
gauge invariant observables that actually vanish in the weak coupling regime
so it can be hoped that correlation functions of these observables are lot
"cleaner". Other important feature of the dual formulation is that it provides
manifestly gauge invariant characterisation of fluctuations.

As shown by \cite{anishetty} there is an intriguing connection between
the dual formulation of Yang-Mills theories in $d=3$ and the Regge-Ponzano
formulation of gravity in $d=3$. A count of the relevant degrees of freedom
indicates that this should happen in $d=4$ also.
\section{The Dual Formulation}
In the conventional formulation, group elements are assigned to the links of
the lattice and gauge invariant action is composed of the traces of the
plaquette variables with the partition sum given by
\beeq
Z = \int\prod_i dU_i e^{\beta \sum_j tr_f P_j} 
\eneq
In order to go over to the dual formulation one introduces the group character expansion
\beeq
exp(\beta tr_f P_i) = \sum_{j_i}(2j_i+1)C_{j_i}(\beta)\chi_{j_i}(P_i)
\eneq
Now either group integrations over products of rotation matrices or graphical
rules \cite{hyper} can be used to carry out the link integrations. In $d=3$,
every link is shared by 4 plaquettes and the link integral is nontrivial if
the representations attached to the four plaquettes are such that $j_1\otimes 
j_2\otimes j_3\otimes j_4$ has a singlet in it. By introducing an auxiliary
$j_5$, this condition can be stated as $(j_1,j_2,j_5),(j_3,j_4,j_5)$ should
satisfy the triangle inequalities. These are the analog of Gauss's law constraints
of the conventional formulation. For lack of space, I'll just quote the final
result \cite{anishetty}
\beeq
Z_{d} =\sum_{\{j\}}\prod (2j+1)C_{j_a}(\beta)
\prod_{i=1}^{5}\left\{\matrix{a_i&b_i&c_i\cr d_i&e_i&f_i\cr }\right\}
\eneq
This partition sum is defined over the dual lattice where $\{j_a\}$ live over
the links and $\{j_b\}$ over the diagonals to the plaquettes. The convention for
the diagonals is that they connect the vertices of the odd sublattice.We have
collectively designated $\{j_a\},\{j_b\}$ by $\{j\}$. Each cube of the dual lattice
is seen to be spanned by 5 tetrahedra of which one is spanned entirely by $\{j_b\}$
while four are spanned by three $\{j_a\}$ and three $\{j_b\}$. Each tetrahedron
carries a weight factor which is the $SU(2)$ 6-j symol $\left\{\matrix{a&b&c\cr
d&e&f\cr}\right\}$. Periodic b.c for the original lattice is crucial for this construction.

The disordered operators $D_i$ satisfy $\chi_{j_i}(P_i)\:\raisebox
{-0.5 ex} {$\stackrel{D_i}{\longrightarrow}$}\: (-1)^{j_i} \chi_{j_i}(P_i)$. 
The disordered 
line is composed of the product of disordered operators for each plaquette
pierced by the disordered line.
\subsection{Connection With Regge-Ponzano Gravity}
The partition function of eqn (3) is the same as the partition function of
the $d=3$ Regge-Ponzano gravity except for the $C_{j_a}$ factors.
These factors break the invariance under the Alexander moves \cite{anishetty}
thereby generating an additional degree of freedom per generator of the 
gauge group. $SU(2)$ gauge theory in $D$ dimensions has $3(D-2)$ d.o.f. The
number of diffeomorphisms being $D$, the d.o.f left is $2D-6$. This correctly
matches the d.o.f of gravity in $D=3,4$ dimensions.
\section{Numerical Simulations}
Apart from the new feature of simulating in the dual variables, there are many
practical advantages also. The basic variables are all integers. Further,
since $C_j(\beta)\:\raisebox{-0.5 ex}{$\stackrel{\beta\rightarrow\infty}
{\longrightarrow}$}\:e^{-{j(j+1)\over 2\beta}}$, only $j_{max}\simeq \sqrt\beta$
is needed making short integers to suffice.
\subsection{Updating}
An allowed configuration is where various triangle inequalities are
satisfied and any updating of variables has to maintain this. Updating
$\{j\}$ independently and rejecting configurations that are not allowed is
wasteful. First we describe what we call a {\bf local} updating algorithm. Each
a-link is common to 4 triangles and each b-link is common to 6 triangles.To
update a link, one first forms $j_{max}^i=j_1^i+j_2^i$ for each triangle $i$
containing the link and where $j_{1,2}$ are the reps on the other two links of
the triangle. Likewise $j_{min}^i=|j_1^i-j_2^i|$ are also calculated and
subsequently $j_{max}=min\{j_{max}^i\}$ and $j_{min}=max\{j_{min}^i\}$. Now the
allowed values for the link to be updated are in the range $(j_{min},j_{max})$
in steps of unity. A particular value can be chosen by either the heatbath
or the Metropolis algorithms.

A major problem with this updating is that it does not change a link from 
integer values(bosonic) to half-integer values(fermionic) and vice versa. Therefore it is not ergodic.
If we think of the integer and half-integer values as a $Z_2$-grading, then these
$Z_2$ d.o.f are not updated by the local method.

Changing only one link from fermionic to bosonic or vice versa is not
possible. In any given triangle at least two such changes must be made. Continuing
this way one is generically led to a proliferation of changes! We must
seek the smallest volume to which such changes can be restricted. If the $Z_2$
flippings are carried out at an odd site, the minimum volume is 8 cubes. These
flippings called {\bf Quasi-local updates} amount to keeping all the links on
the outer surface fixed and flipping all the interrior links. If on the other
hand the flippings are done at an even site, the minimum volume is a cluster
of 8 tetrahedra.

One still faces the question of how these flippings are to be done satisfying all
the triangle inequalities. That brings us to the use of Kagome(K) variables \cite
{gadiyar}. The basic idea is that the triangle inequalities can be traded for
{\it equalities}. If $(j_1,j_2,j_3)$ are the links of a triangle, introducing the
variables $j_1=n_2+n_3,j_2=n_3+n_1,j_3=n_1+n_2$, it can be shown that choosing
$n_i$ freely among positive integers, one can generate all the allowed $j_i$. But
for our case, if a link is shared between two triangles, then the sum of 
K-variables for this link from one triangle must equal the sum of the corresponding
K-variables for the other triangle. Thus K-variables are not all
{\it independent}.

Now the solution to the quasi-local update problem in terms of the K-variables
is as follows: i) determine the K-variables for the links on the outer
surface and keep them fixed; ii) determine the {\it the set of independent 
K-variables} for the interrior; iii) change the interrior variables freely subject
to realising the fermion-boson flippings for the interrior links. A {\it simplified}
quasi-local update consists of simply shifting all $\{j\}$ by $\{j\pm 1/2\}$.

The $Z_2$-fluctuations can be controlled by switching on/off the quasi-local
updates at any site one wishes. Thermalisation can be achieved first with the
local updates and on the resulting ensemble one can perform the quasi-local
updates till the $Z_2$-configurations thermalise. In this manner the precise
influence of the $Z_2$-fluctuations on various dynamical mechanisms can be studied.

Finally it is worth mentioning as to how best the 6-j evaluations should be done.
Calculating them for each configuration is unnecessarily time consuming. For local
updates it is possible to use a recursive algorithm. But it is best to make 
a table of only the non-vanishing 6-j's
into an one-dimensional array with an unique labelling.
\subsection{Positivity of Weights?}
The product of 6-j symbols is generally not of a particular sign. Nevertheless, 
based on my experiences with the $d=4$ hypercube problem \cite{hyper} it was
initially hoped that a configuration of positive weight remains so under all
updates. Over a large number of sweeps this was indeed so, but eventually negative
weight configurations started appearing. Barring a phase error in the analytical
calculations, it appeared as if the dual formulation had seen the end of the line
at least as far as numerical simulations were concerned. 

But a resolution of this problem has recently occurred to me. Let us say the
weight factors $p_1,p_2,p_3,....,p_k,...$ are such that some of them, say,
$p_l,p_m,p_n,...$ are negative. Now the trick is to generate configurations
with the probabilities $\{\tilde p_i={|p_i|\over P}\}$ where $P=\sum_i |p_i|$ but keep
track of the configurations with negative weight through $\epsilon_i$ which
is $+1$ when $p_i > 0$ and $-1$ when $p_i < 0$.The expectation value of an 
observable with value $O_i$ in the configuration $i$ is now given by
\beeq
\langle O \rangle_p = {\langle O\epsilon \rangle_{\tilde p}\over \langle \epsilon \rangle_{\tilde p}}
\eneq
Since the partition sum $\langle 1 \rangle_p = \langle \epsilon\rangle_{\tilde p}\ne 0$, this procedure is well 
defined.
\section{Open Issues}
Among the important open problems are the generalisation of our algorithms to $d=4$
using the results of \cite{halliday}. It would also be interesting to generalise
our results for the case of the models considered by Ooguri\cite{ooguri} and their 
non-topological versions. Inclusion of matter particularly fermions is another
outstanding problem. Since we are eventually interested only in the large $\beta$
regime a simpler version using only the asymptotic forms of the 6-j symbols may
be possible. Generalisation to $SU(3)$ is another open direction. Finally mention
must be made of what I call {\it exceptional configurations} for which some of
the 6-j's vanish even though all the triangle inequalities are satisfied. There
are fascinating group theoretical explanations for these. Their contribution to
the partition sum is zero. One may speculate as to the physical significance of
these configurations for gauge theories.

\end{document}